\documentclass[smallextended]{svjour3}

\smartqed  
\usepackage{graphicx}
\usepackage{array}
\usepackage{cite}
\usepackage{float}
\usepackage{subfigure}
\usepackage{amsmath,amssymb,mathrsfs}
\newtheorem{theo}{Theorem}

\journalname{Advances in Data Analysis and Classification}
\begin{document}
\title{A Global Algorithm for Clustering Univariate Observations}

\author{Nicolas Paul \and
        Michel Terre \and
        Luc Fety 
}

\institute{N. Paul, M. Terre, L. Fety \at
              Conservatoire National des Arts et Metiers, Electronic and Communications, \\
              292 rue Saint-Martin, 75003 PARIS, FRANCE \\
              Tel.: 33 1 40 27 25 67\\
              Fax: 33 1 40 27 24 81\\
              \email{nicolas.paul@cnam.fr}   }

\date{Received: date / Accepted: date}

\maketitle

\begin{abstract}
This paper deals with the clustering of univariate observations: given a set of observations coming from $K$ possible clusters, one has to estimate the cluster means. We propose an algorithm based on the minimization of the "KP" criterion we introduced in a previous work. In this paper, we show that the global minimum of this criterion can be reached by first solving a linear system then calculating the roots of some polynomial of order $K$. The KP global minimum provides a first raw estimate of the cluster means, and a final clustering step enables to recover the cluster means. Our method's relevance and superiority to the Expectation-Maximization algorithm is illustrated through simulations of various Gaussian mixtures. 
\end{abstract}

\section{Introduction}
In this paper we focus on the clustering of univariate observations coming from $K$ possible clusters, when the number of clusters is known. One method consists in estimating the observation pdf (mixture of $K$ pdf), by associating a kernel to each observation and adding the contribution of all the kernels (Parzen 1962). A search of the pdf modes then leads to the cluster means. The drawback of such method is that it requires the configuration of extra-parameters (kernel design, intervals for the mode search). Alternately, the Expectation-Maximization (EM) (Dempster et al. 1977) algorithm is the most commonly used method when the mixture densities belong to the same known parameterized family. It is an iterative algorithm that look for the mixture parameters that maximize the likelihood of the observations. Each EM iteration consists of two steps. The Expectation step estimates the probability for each observation to come from each mixture component. Then, during the Maximization step, these estimated probabilities are used to update the estimation of the mixture parameters. One can show that this procedure converges to one maximum (local or global) of the likelihood (Dempster et al. 1977). If the mixture components do not belong to a common and known parameterized family, the EM algorithm does not directly apply. Yet, if the component densities do not overlap too much, some clustering methods can be used to cluster the data and calculate the cluster means: in (Fisher 1958) an algorithm is proposed to compute the $K$-partition of the $N$ sorted observations which minimize the sum of the squares within clusters. Instead of testing the $\binom{N-1}{K-1}$ possible partitions, some relationships between $k$-partitions and $(k+1)$-partitions are used to recursively compute the optimal $K$-partition. The main drawbacks of this method are a high sensitivity to potential differences between the cluster variances and a complexity in $\text{O}(KN^2)$ (Fitzgibbon 2000). Among the clustering methods, the K-Means algorithm (Hartigan 1977) is one of the most popular method. It is an iterative algorithm which groups the data into K clusters in order to minimize an objective function such as the sum of point to cluster mean square Euclidean distance. The main drawback of K-Means or EM is the potential convergence to some local extrema of the criterion they use. Some solutions consist for instance in using smart initializations (see (McLachlan and Peel 2000) and (Lindsay and Furman 1994) for EM, (Bradley and Fayyad 1998) for k-means) or stochastic optimization, to become less sensitive in the initialization (see (Celeux et al. 1995) and (Pernkopf and Bouchaffra 2005) for EM, (Krishna and Murty 1999) for K-means). Another drawback of these methods is the convergence speed, which can be very slow when the number of observations is high. A survey of the clustering techniques can be found in (Berkhin 2006) and (Xu and Wunsch 2005). In this contribution, we propose a non-iterative algorithm which mainly consists in calculating the minimum of the "K-Product" (KP) criterion we first introduced in (Paul et al. 2006): if $\{z_n\}_{n\in\{1 \cdots N\}}$ is a set of $N$ observations, $K$ the known number of clusters and $\{x_k\}_{k\in\{1 \cdots K\}}$ any vector of $\mathbb{R}^K$, we define the KP criterion as the sum of all the K-terms products $\prod_{k=1}^{K}(z_n-x_k)^2$. The main motivation for using such criterion is that, though it provides a slightly biased estimation of the cluster means, its global minimum can be reached by first solving a linear system then calculating the roots of some polynomial of order $K$. Once these $K$ roots have been obtained, a final clustering step assigns each observation to the closest root and calculates the resulting cluster means. Another advantage of the proposed method is that it does not require the configuration of any extra-parameters. The rest of the paper is organized as follow: In section 2 the observation model is presented and the criterion is defined. In section 3 the criterion global minimum is theoretically calculated. In section 4 the clusters estimation algorithm is described. Section 5 presents simulation results which illustrate the algorithm performances on different Gaussian mixtures: mixtures of three, six and nine components have been simulated with various configurations (common/different mixing weights, common/different variances). Conclusions are finally given in Section 6.
 
\section{Observation model and criterion definition}
Let $\{ a_k \}_{k\in\{1 \cdots K\}}$ be a set of $K$ different values of $\mathbb{R}^K$, let $\textbf{a}$ be the vector of cluster means defined by $\textbf{a}\stackrel{\Delta}{=}(a_1,a_2 \cdots a_K)^t$, let $\{ \pi_k \}_{k\in\{1 \cdots K\}}$ be a set of $K$ mixing weights (prior probabilities) that sum up to one and let $\{g_k\}_{k\in\{1 \cdots K\}}$ be a set of $K$ zeros-mean densities. The probability density function of the multimodal observation $z$ is a finite mixture given by: 
\begin{equation}
f(z)=\sum_{k=1}^{K}{ \pi_k g_k(z-a_k) } \notag
\end{equation}
\noindent Note that the form of the densities $g_k$ are usually not known by the estimator and that the $g_k$ do not necessarily belong to the same parameterized family. Now let $\{ z_n \}_{n\in\{1 \cdots N\}}$ be a set of $N$ observations in $\mathbb{R}^N$. In all the following we assume that $N$ is greater than $K$ and that the number of different observations is greater than $K-1$. The KP criterion $J(\textbf{x})$ is defined by:
\begin{equation}
J: \mathbb{R}^K\rightarrow \mathbb{R}^+: \  
\textbf{x} \rightarrow \sum_{n=1}^{N}{ \prod_{k=1}^{K}{\left(z_n-x_k\right)^2} }
\label{definition_J} 
\end{equation}
\noindent Note the difference with the K-means criterion which can be written (for the square Euclidean distance):
\begin{equation}
\text{K-means}: \mathbb{R}^K\rightarrow \mathbb{R}^+: \  
\textbf{x} \rightarrow 
\sum_{n=1}^{N}{ \underset{k\in\{1 \cdots K\}}{\text{min}} (z_n-x_k)^2 } \notag
\end{equation}
\noindent The KP criterion \eqref{definition_J} is clearly positive for any vector $\textbf{x}$. The first intuitive motivation for defining this criterion is its asymptotic behavior when all the $g_k$ variances are null. In this case, all the observations are equal to one of the $a_k$ and therefore $J(\textbf{a})=0$. $J(\textbf{x})$ is then minimal when $\textbf{x}$ is equal to $\textbf{a}$ or any of its $K!$ permutations. The second motivation is that, in the general case, $J$ have $K!$ minima that are the $K!$ permutations of one single vector which can be reached by solving a linear system then finding the roots of some polynomial of order $K$. This is shown in section 3.

\section{KP global minimum}
We first give in section \ref{section3A} some useful definitions which are needed in section \ref{section3B} to reach the global minimum of $J$.
\subsection{Some useful definitions}
\label{section3A}
To any observation $z_n$ we associate the vector $\textbf{z}_n$ defined by:
\begin{equation}
\textbf{z}_n\stackrel{\Delta}{=}(z_n^{K-1}, z_n^{K-2} \cdots ,1)^t, \ \ \textbf{z}_n \in \mathbb{R}^K
\label{definition_zn} 
\end{equation}
\noindent The vector $\textbf{z}$ and the Hankel matrix $\textbf{Z}$ are then respectively defined by:
\begin{equation}
\textbf{z}\stackrel{\Delta}{=}\sum_{n=1}^{N}{z_n^K \textbf{z}_n}, \ \ \textbf{z} \in \mathbb{R}^K
\label{definition_z} 
\end{equation}
\begin{equation}
\textbf{Z}\stackrel{\Delta}{=}\sum_{n=1}^{N}{ \textbf{z}_n \textbf{z}_n^t }, \ \ \textbf{Z} \in \mathbb{R}^{K \times K}
\label{definition_Z} 
\end{equation}
\noindent The matrix \textbf{Z} is regular if the number of different observations is greater than $K-1$ (one explanation is detailed in Appendix A).\\
\\
\noindent Now let $\textbf{y}=(y_1,\cdots,y_K)^t$ be a vector of $\mathbb{R}^K$. We define the polynomial of order $K$ $q_\textbf{y}(\alpha)$ as:
\begin{equation}
q_{\textbf{y}}(\alpha)\stackrel{\Delta}{=} \alpha^K-\sum_{k=1}^{K}{\alpha^{K-k}y_k}
\label{definition_qy} 
\end{equation}
\noindent if $\textbf{r}=(r_1,\cdots,r_K)^t$ is a vector of $\mathbb{C}^K$ containing the $K$ roots of $q_{\textbf{y}}(\alpha)$ the factorial form of $q_{\textbf{y}}(\alpha)$ is:
\begin{equation}
q_{\textbf{y}}(\alpha)=\prod_{k=1}^{K}(\alpha-r_k) \notag
\end{equation}
\begin{equation}
q_{\textbf{y}}(\alpha)=\alpha^K-(r_1+\cdots+r_K)\alpha^{K-1}+...\\
+(-1)^K(r_1\times r_2 \cdots \times r_K) \notag
\end{equation}
\begin{equation}
q_{\textbf{y}}(\alpha)=\alpha^K-\sum_{k=1}^{K}{\alpha^{K-k}w_k(\textbf{r})} \notag
\end{equation}
\noindent where $w_k(\textbf{r})$ is the Elementary Symmetric Polynomial (ESP) in the variables ${r_1,\cdots,r_K}$ defined by:
\begin{equation}
w_k(\textbf{r}) \stackrel{\Delta}{=} (-1)^{k+1} \sum_{ \substack{ \{j_1,\cdots,j_k\} \in \{1\cdots K\}^k \\ j_1<\cdots<j_k \leqslant K } }{r_{j_1}.r_{j_2}\cdots.r_{j_k}}
\label{definition_wk} 
\end{equation}
\noindent For instance, for $K=3$, we have:
\begin{equation}
w_1(\textbf{r})=r_1+r_2+r_3 \notag
\end{equation}
\begin{equation}
w_2(\textbf{r})=-(r_1r_2+r_2r_3+r_1r_3) \notag
\end{equation}
\begin{equation}
w_3(\textbf{r})=r_1r_2r_3 \notag
\end{equation}
\noindent If we call $\textbf{w}(\textbf{r})$ the vector of ESP of $\textbf{r}$ defined by:
\begin{equation}
\textbf{w}(\textbf{r})\stackrel{\Delta}{=}(w_1(\textbf{r}),\cdots,w_K(\textbf{r}))^t
\label{definition_w} 
\end{equation}
\noindent the relationship between the roots and coefficients of $q_{\textbf{y}}(\alpha)$ becomes:
\begin{equation}
\textbf{y}=\textbf{w}(\textbf{r}) \Leftrightarrow \forall k \in \{1\cdots K\} \ q_{\textbf{y}}(r_k)=0  
\label{roots_coefficients} 
\end{equation}

\subsection{The KP minimum}
\label{section3B}
The global minimum of $J$ is given by theorem 1:
\begin{theo}
if $\textbf{y}_{min}$ is the solution of $\textbf{Z}.\textbf{y}_{min}=\textbf{z}$ (where $\textbf{z}$ and $\textbf{Z}$ have been defined in \eqref{definition_z} and \eqref{definition_Z}) and if $\textbf{x}_{min}$ is a vector containing, in any order, the $K$ roots of $q_{\textbf{y}_{min}}(\alpha)$ (defined in \eqref{definition_qy}), then $\textbf{x}_{min}$ belongs to $\mathbb{R}^K$ and $\textbf{x}_{min}$ is the global minimum of $\textbf{J}$. 
\end{theo}
The proof is given in appendix B.

\section{Clusters estimation algorithm}
The clusters estimation algorithm consists of two steps. In the first step, the minimum of $J$, $\textbf{x}_{min}=(x_{1,min},...,x_{K,min})^t$, is calculated, giving a first raw estimation of the set of cluster means. This first estimate is slighly biased: for instance, for a Gaussian mixture with two balanced components centred on $-a$ and $a$ and a common standard deviation $\sigma$ the asymptotical solution of $\textbf{Z}.\textbf{y}_{min}=\textbf{z}$ is $\textbf{y}_{min}=(0,a^2+\sigma^2)^t$ and the roots of $q_{\textbf{y}_{min}}(\alpha)$ are:
\begin{equation}
\textbf{x}_{min}=\left(-a\sqrt{1+\frac{\sigma^2}{a^2}},a\sqrt{1+\frac{\sigma^2}{a^2}}\right) \notag
\end{equation}
\noindent Therefore, in a second step, each observation $z_n$ is assigned to the nearest $x_{k,min}$, $K$ clusters are formed, and the final estimated cluster means are calculated. The algorithm steps and their complexities are illustrated in table~\ref{algo}. The total complexity is in O$(NK+K^2)$, which is equivalent to O$(NK)$ since $N$ is greater than $K$. 
\begin{table}
\renewcommand{\arraystretch}{1.3}
\caption{KP algorithm steps and complexities}
\label{algo}
\centering
\begin{tabular} {c}
\hline
\bfseries step 1: calculate a minimum of J \\
\hline
calculate $\textbf{Z}$ and $\textbf{z}$: O$(NK)$ \\
calculate $\textbf{y}_{min}$ by solving $\textbf{Z}.\textbf{y}_{min}=\textbf{z}$: O$(K^2)$ \\
calculate the roots 
$(x_{1,min}, \cdots ,x_{K,min})$ 
of $q_{\textbf{y}_{min}}(\alpha)$: O$(K^2)$ \\
\hline
\bfseries step 2: clustering and cluster means estimation \\
\hline
assign each $z_n$ to the closest $x_{k,min}$: O$(NK)$ \\
calculate the K means of the resulting clusters: O$(N)$ \\
\hline
\end{tabular}
\end{table}

\section{Simulations}
Several types of Gaussian mixture have been considered. The number of components (clusters) is equal to three (scenario A), six (scenario B) and nine (scenario C). In scenario A, the distance between two successive cluster centers (component means) is equal to one. In scenario B and in scenario C, the distance between two successive centers is equal to one or two. For each scenario "X", four cases have been studied: common variance and common mixing weight (scenario X.1), different variances and common mixing weight (scenario X.2), common variance and different mixing weights (scenario X.3) and different variances and different mixing weights (scenario X.4). A summary of all the scenari is given in Tables 2, 3 and 4. The number of observations ($N$) per simulation run is equal to 100 in scenario A, 200 in scenario B and 300 in scenario C.

\begin{table}
\caption{simulation scenario A}
\begin{center}{
\begin{tabular}{|c|c c|c c|c c|c c|}
\cline{2-9}
\multicolumn{1}{c|}{} & \multicolumn{2}{c|}{scenario A.1} & \multicolumn{2}{c|}{scenario A.2} & \multicolumn{2}{c|}{scenario A.3} & \multicolumn{2}{c|}{scenario A.4} \\[2mm]
\hline
means & variances & prior & variances & prior & variances & prior & variances & prior \\[2mm]
\hline
$0$ & $\sigma^2$ & $\frac{1}{3}$ & $\sigma^2$ & $\frac{1}{3}$ & $\sigma^2$ & $0.4$ & $\sigma^2$ & $0.4$ \\[2mm]
\hline
$1$ & $\sigma^2$ & $\frac{1}{3}$ & $\frac{\sigma^2}{2}$ & $\frac{1}{3}$ & $\sigma^2$ & $0.4$ & $\frac{\sigma^2}{2}$ & $0.4$ \\[2mm]
\hline
$2$ & $\sigma^2$ & $\frac{1}{3}$ & $\sigma^2$ & $\frac{1}{3}$ & $\sigma^2$ & $0.2$ & $\sigma^2$ & $0.2$ \\[2mm]
\hline
\end{tabular}}
\end{center}
\end{table}

\begin{table}
\caption{simulation scenario B}
\begin{center}{
\begin{tabular}{|c|c c|c c|c c|c c|}
\cline{2-9}
\multicolumn{1}{c|}{} & \multicolumn{2}{c|}{scenario B.1} & \multicolumn{2}{c|}{scenario B.2} & \multicolumn{2}{c|}{scenario B.3} & \multicolumn{2}{c|}{scenario B.4} \\[2mm]
\hline
means & variances & prior & variances & prior & variances & prior & variances & prior \\[2mm]
\hline
$0$ & $\sigma^2$ & $\frac{1}{6}$ & $\sigma^2$ & $\frac{1}{6}$ & $\sigma^2$ & $0.2$ & $\sigma^2$ & $0.2$ \\[2mm]
\hline
$1$ & $\sigma^2$ & $\frac{1}{6}$ & $\frac{\sigma^2}{2}$ & $\frac{1}{6}$ & $\sigma^2$ & $0.2$ & $\frac{\sigma^2}{2}$ & $0.2$ \\[2mm]
\hline
$2$ & $\sigma^2$ & $\frac{1}{6}$ & $\sigma^2$ & $\frac{1}{6}$ & $\sigma^2$ & $0.1$ & $\sigma^2$ & $0.1$ \\[2mm]
\hline
$4$ & $\sigma^2$ & $\frac{1}{6}$ & $\frac{\sigma^2}{2}$ & $\frac{1}{6}$ & $\sigma^2$ & $0.2$ & $\frac{\sigma^2}{2}$ & $0.2$ \\[2mm]
\hline
$5$ & $\sigma^2$ & $\frac{1}{6}$ & $\sigma^2$ & $\frac{1}{6}$ & $\sigma^2$ & $0.2$ & $\sigma^2$ & $0.2$ \\[2mm]
\hline
$6$ & $\sigma^2$ & $\frac{1}{6}$ & $\frac{\sigma^2}{2}$ & $\frac{1}{6}$ & $\sigma^2$ & $0.1$ & $\frac{\sigma^2}{2}$ & $0.1$ \\[2mm]
\hline
\end{tabular}}
\end{center}
\end{table}

\begin{table}
\caption{simulation scenario C}
\begin{center}{
\begin{tabular}{|c|c c|c c|c c|c c|}
\cline{2-9}
\multicolumn{1}{c|}{} & \multicolumn{2}{c|}{scenario C.1} & \multicolumn{2}{c|}{scenario C.2} & \multicolumn{2}{c|}{scenario C.3} & \multicolumn{2}{c|}{scenario C.4} \\[2mm]
\hline
means & variances & prior & variances & prior & variances & prior & variances & prior \\[2mm]
\hline
$0$ & $\sigma^2$ & $\frac{1}{9}$ & $\sigma^2$ & $\frac{1}{9}$ & $\sigma^2$ & $\frac{2}{15}$ & $\sigma^2$ & $\frac{2}{15}$ \\[2mm]
\hline
$1$ & $\sigma^2$ & $\frac{1}{9}$ & $\frac{\sigma^2}{2}$ & $\frac{1}{9}$ & $\sigma^2$ & $\frac{2}{15}$ & $\frac{\sigma^2}{2}$ & $\frac{2}{15}$ \\[2mm]
\hline
$2$ & $\sigma^2$ & $\frac{1}{9}$ & $\sigma^2$ & $\frac{1}{9}$ & $\sigma^2$ & $\frac{1}{15}$ & $\sigma^2$ & $\frac{1}{15}$ \\[2mm]
\hline
$4$ & $\sigma^2$ & $\frac{1}{9}$ & $\sigma^2$ & $\frac{1}{9}$ & $\sigma^2$ & $\frac{1}{15}$ & $\sigma^2$ & $\frac{1}{15}$ \\[2mm]
\hline
$5$ & $\sigma^2$ & $\frac{1}{9}$ & $\frac{\sigma^2}{2}$ & $\frac{1}{9}$ & $\sigma^2$ & $\frac{3}{15}$ & $\frac{\sigma^2}{2}$ & $\frac{3}{15}$ \\[2mm]
\hline
$6$ & $\sigma^2$ & $\frac{1}{9}$ & $\sigma^2$ & $\frac{1}{9}$ & $\sigma^2$ & $\frac{1}{15}$ & $\sigma^2$ & $\frac{1}{15}$ \\[2mm]
\hline
$8$ & $\sigma^2$ & $\frac{1}{9}$ & $\sigma^2$ & $\frac{1}{9}$ & $\sigma^2$ & $\frac{2}{15}$ & $\sigma^2$ & $\frac{2}{15}$ \\[2mm]
\hline
$9$ & $\sigma^2$ & $\frac{1}{9}$ & $\frac{\sigma^2}{2}$ & $\frac{1}{9}$ & $\sigma^2$ & $\frac{2}{15}$ & $\frac{\sigma^2}{2}$ & $\frac{2}{15}$ \\[2mm]
\hline
$10$ & $\sigma^2$ & $\frac{1}{9}$ & $\sigma^2$ & $\frac{1}{9}$ & $\sigma^2$ & $\frac{1}{15}$ & $\sigma^2$ & $\frac{1}{15}$ \\[2mm]
\hline
\end{tabular}}
\end{center}
\end{table}


\noindent To evaluate the performances of our proposal we compare it to the classical EM algorithm. To estimate the parameters of a Gaussian mixtures, the EM algorithm proceeds as follows (Dempster et al. 1977): if $\hat{\beta}^{(ite)}_{n,k}$ is the estimed probability that $z_n$ comes from cluster $k$ and if $\hat{\pi}^{(ite)}_k$, $\hat{a}^{(ite)}_k$ and $\hat{\sigma}^{(ite)}_k$ are respectively the estimated prior, mean and standard deviation of cluster $k$ at iteration $ite$, then the estimations at iteration $ite+1$ are given by: \\
\noindent Expectation step:
\begin{equation}
\hat{\beta}^{(ite+1)}_{n,k}=
\frac{\frac{\hat{\pi}^{(ite)}_k}{\sqrt{2\pi}\hat{\sigma}^{(ite)}_k} 
\text{exp}\left(-\frac{1}{2}\left(\frac{z_n-\hat{a}^{(ite)}_k}{\hat{\sigma}^{(ite)}_k}\right)^2\right)}
{\sum\limits_{k=1}^{K}{ \frac{\hat{\pi}^{(ite)}_k}{\sqrt{2\pi}\hat{\sigma}^{(ite)}_k} \text{exp}\left(-\frac{1}{2}\left(\frac{z_n-\hat{a}^{(ite)}_k}{\hat{\sigma}^{(ite)}_k}\right)^2\right)}} \notag
\end{equation}
\noindent Maximization step:
\begin{equation}
\hat{\pi}^{(ite+1)}_k=\frac{ \sum_{n=1}^{N}{ \hat{\beta}^{(ite+1)}_{n,k}} }{N} \notag                          
\end{equation}
\begin{equation}
\hat{a}^{(ite+1)}_k=\frac{ \sum_{n=1}^{N}{ \hat{\beta}^{(ite+1)}_{n,k} }z_n }{ \sum_{n=1}^{N}{ \hat{\beta}^{(ite+1)}_{n,k} }} \notag
\end{equation}
\begin{equation}
\hat{\sigma}^{(ite+1)}_k=\frac{ \sum_{n=1}^{N}{\hat{\beta}^{(ite+1)}_{n,k}}\left(z_n-\hat{a}^{(ite+1)}_k\right)^2 }{ \sum_{n=1}^{N}{\hat{\beta}^{(ite+1)}_{n,k}}} \notag
\end{equation}
This iterative procedure converges to one maximum of the likelihood function $ \prod\limits_{n=1}^{N}{ P(z_n | \left\{ \hat{a}_k \hat{\sigma}_k \hat{\pi}_k \right\}_{k\in \{1 \cdots K\}})}$. To initialize the EM algorithm in our simulations, $K$ cluster means $\hat{a}^{(0)}_k$ are randomly chosen with a uniform draw in the observation zone $[\text{min}(z_n) \quad \text{max}(z_n)]$. For each $n$, $\hat{\beta}^{(1)}_{n,k}$ is set to one if $\hat{a}^{(0)}_k$ is the closest cluster means to the observation $z_n$ and $\hat{\beta}^{(1)}_{n,k}$ is set to zero otherwise. This initialization is repeated until each cluster contains at least one observation. Then the EM starts with a maximization step. The algorithm is stopped if all the estimated parameters do not change between two EM steps or if a maximal number of $100$ iterations is reached.

The clustering performances are evaluated as follows: to get rid of the permutation ambiguity, for each simulation run $r$ and estimation $\hat{\textbf{a}}_r$, the performance criterion $e_r$ is defined as the maximal absolute distance between the true and estimated sorted vector of cluster means: 
\begin{equation}
e_r\stackrel{\Delta}{=}\text{N} \left( \text{sort}(\textbf{a})-\text{sort}(\hat{\textbf{a}}_r)\right) \notag
\end{equation}
where $\text{N}(\textbf{x})\stackrel{\Delta}{=} \underset{k \in \{1\cdots K\}}{\text{max}} |x_k|$. 

The distribution of $e_r$ is given in figure 1 for the scenario A.1 with $\sigma=0.25$ and 10000 simulation run. The KP minimum is a biased estimation: $e_r$ is greater than $0.1$ for $90\%$ of the run. Yet, $e_r$ remains less than $0.2$ for $80\%$ of the run. Then the "full KP" algorithm (calculation of the KP minimum followed by a clustering) always provides an accurate set of estimates: $e_r$ remains less than $0.1$ (resp $0.2$) for $80\%$ (resp. $100\%$) of the run. With the EM algorithm, $e_r$ is less than $0.1$ (resp. $0.2$) for $45\%$ (resp. $65\%$) of the run but $e_r$ is greater than $0.5$ for $30\%$ of the run. In this case, the EM gets stuck at a local maximum of the likelihood. Typically one estimated cluster mean is located in the middle of two true cluster means (assuming a too high variance), while two other estimated cluster means are closed to the same true cluster mean. In figure 2, 3 and 4 we present the EM and KP performances for all the scenari with different values of $\sigma$. In scenario A (scenari A.1 to A.4) the KP algorithm estimation is perfect when $\sigma$ is less than $0.2$ ($e_r$ is less than $0.1$ for $95\%$ of the run) and remains correct for $\sigma<0.3$ ($e_r$ is less than $0.2$ for $95\%$ of the run). On the contrary, EM can provide a wrong set of estimated clusters as soon as $\sigma$ is not null: for instance, when $\sigma=0.1$, $e_r$ is greater than $0.5$ for $25\%$ of the run. When the mixture components strongly overlap ($\sigma>0.5$) the two methods lead to wrong estimations, with a slight superiority of EM when $\sigma>0.8$. The KP algorithm remains superior to EM in scenario B and in scenario C. In scenario B (6 clusters), KP is robust for any $\sigma$ less than $0.15$, while, for $\sigma=0.1$, EM converges to a wrong set of cluster means for $75\%$ of the run. in scenario C (9 clusters), KP is robust for any $\sigma$ less than 0.05, while, for $\sigma=0.02$, EM converges to a wrong set of cluster means for $70\%$ of the run. In each scenario, the mixing weights configuration (balanced/unbalanced) has a slight influence on the KP algorithm: the performances on sub-scenari X.3 and X.4 (different mixing weights) are weaker than the performances on sub-scenari X.1 and X.2 (common mixing weights). Yet the KP performances on the unbalanced mixtures remain strongly greater than the EM performances.

\begin{figure}
\begin{center}{
\resizebox{0.75\textwidth}{!}{%
  \includegraphics{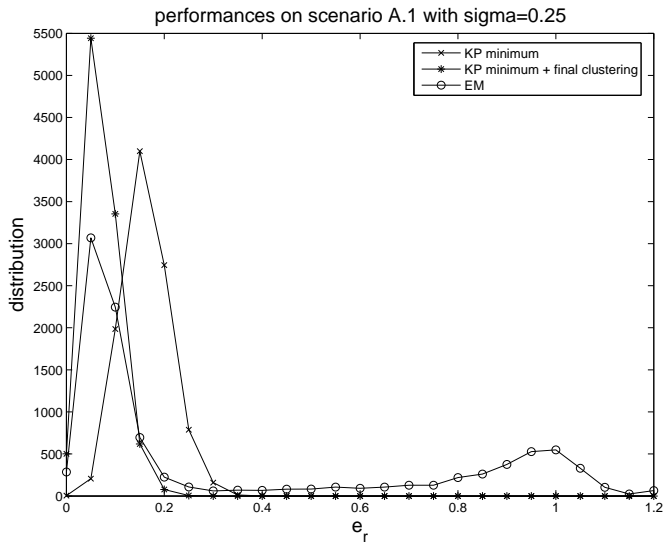}
}}
\end{center}
\caption{Estimation performances on scenario A.1 with $\sigma=0.25$. 10000 simulation run have been performed. For each simulation run, 100 observations have been generated. $e_r$ is the maximal distance between the sorted vector of true cluster means and the sorted vector of estimated cluster means}
\label{fig:1}       
\end{figure}

\begin{figure}{
\begin{center}{
\begin{subfigure}{
\includegraphics[height=0.45\linewidth]{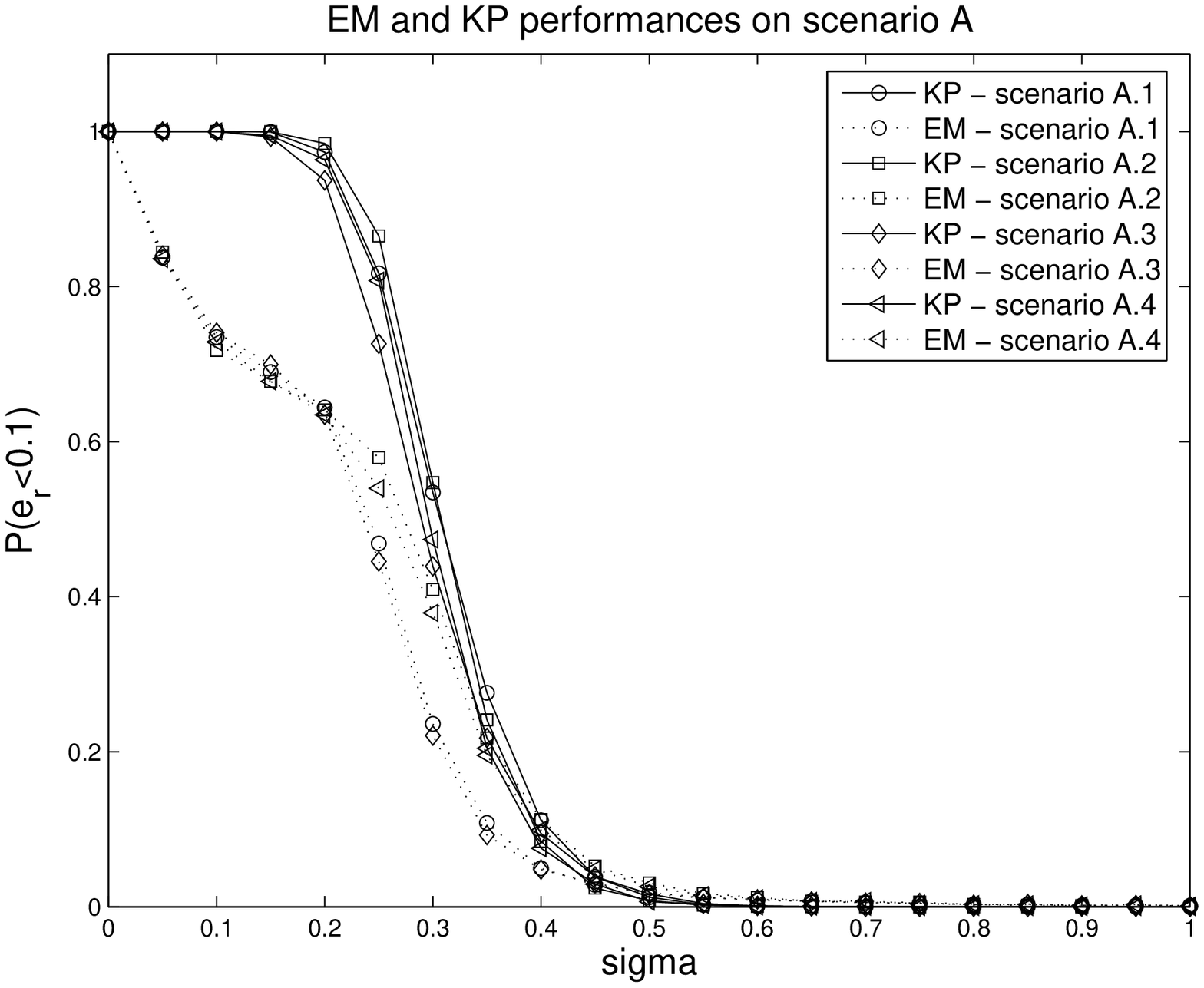}}
\end {subfigure}
\begin{subfigure}{
\includegraphics[height=0.45\linewidth]{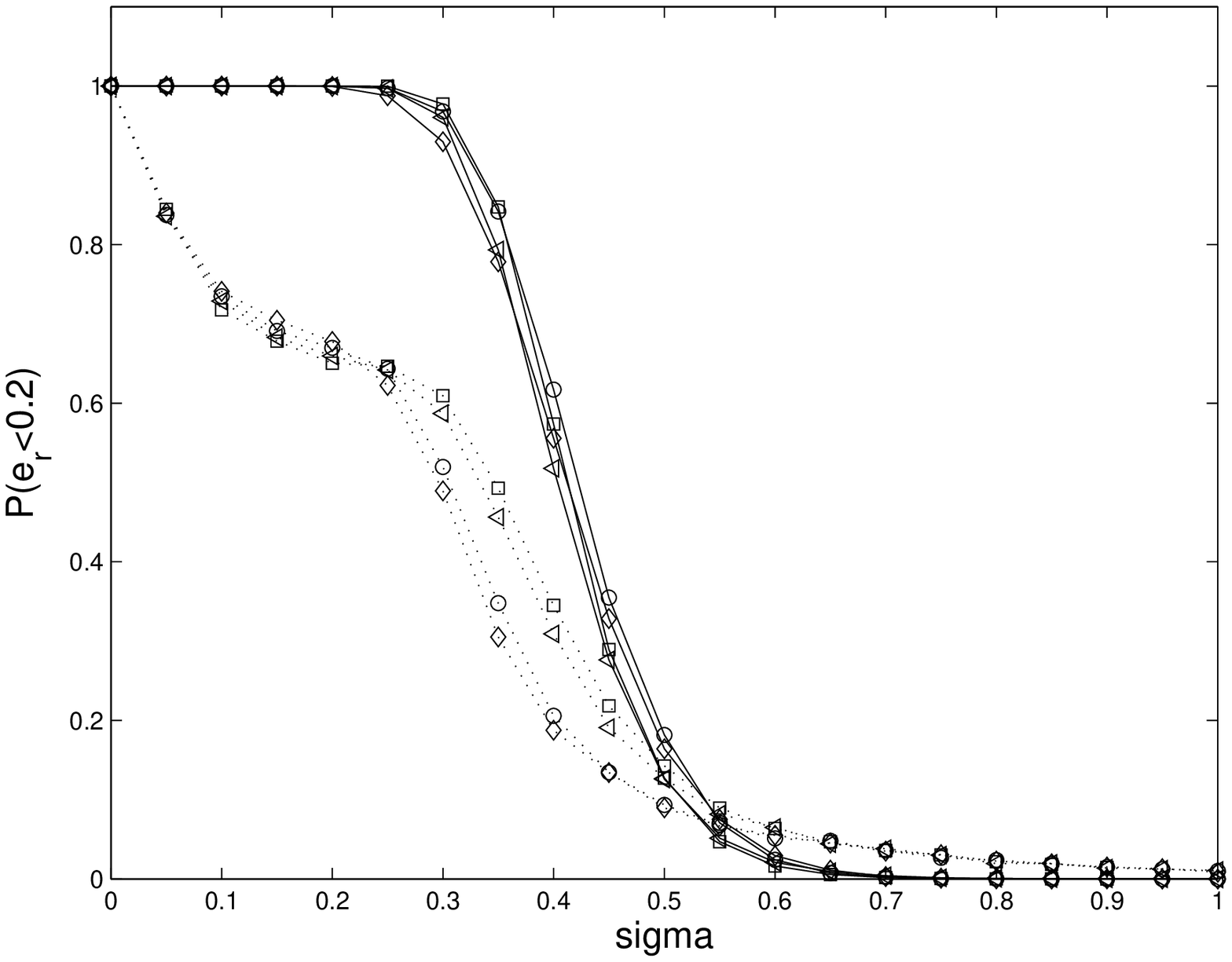}}
\end{subfigure}
\begin{subfigure}{
\includegraphics[height=0.45\linewidth]{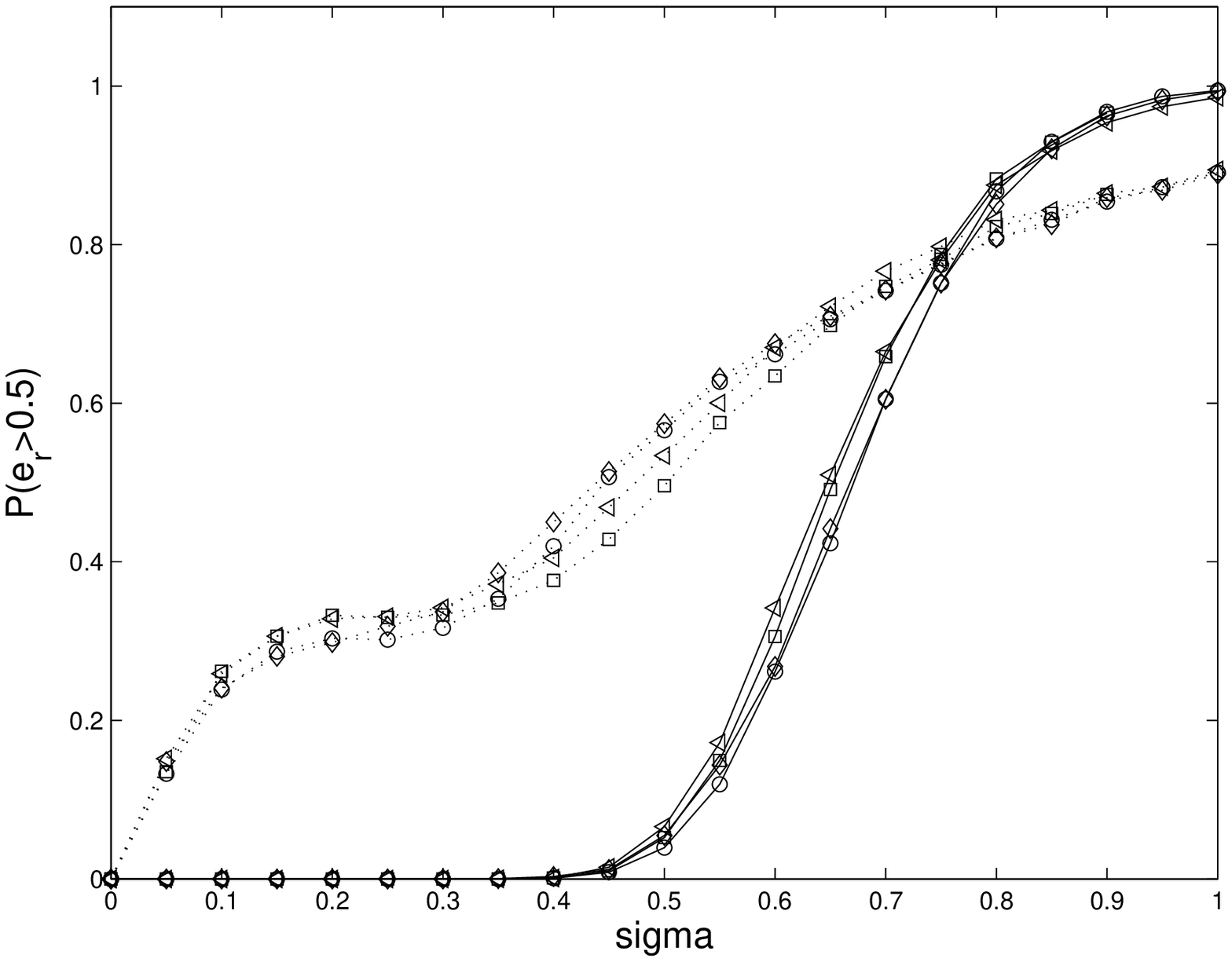}}
\end{subfigure}}
\end{center}}
\caption{performances of the EM and KP algorithms on scenario A for different values of $\sigma$. For each value of $\sigma$ and for each sub-scenario 10000 simulation run have been performed. For each simulation run, 100 observations have been generated. $e_r$ is the maximal distance between the sorted vector of true cluster means and the sorted vector of estimated cluster means. The performance criteria are the probabilities for $e_r$ to be smaller than 0.1 (top), smaller than 0.2 (middle) and greater than 0.5 (bottom).} 
\end{figure}

\begin{figure}{
\begin{center}{
\begin{subfigure}{
\includegraphics[height=0.45\linewidth]{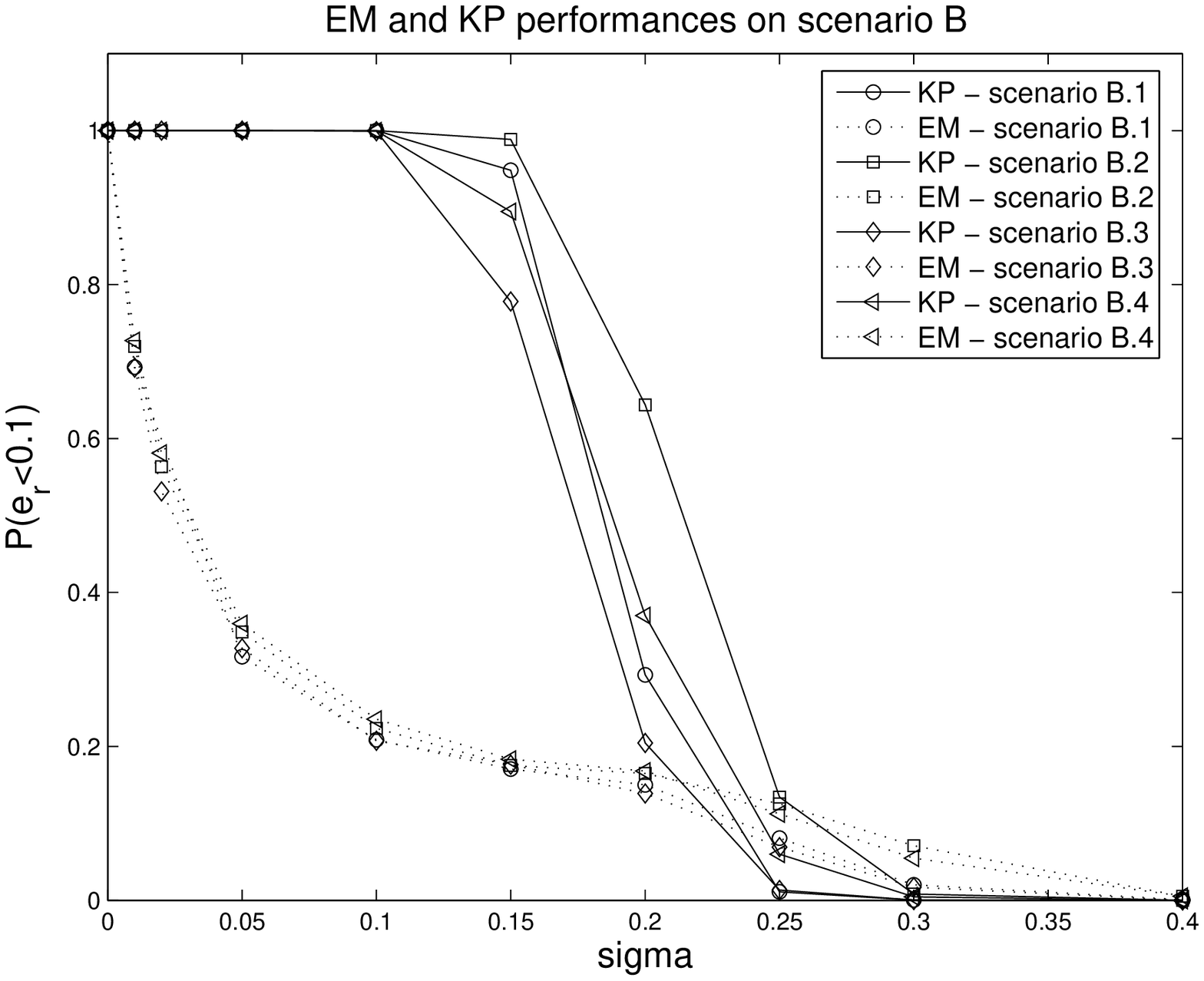}}
\end {subfigure}
\begin{subfigure}{
\includegraphics[height=0.45\linewidth]{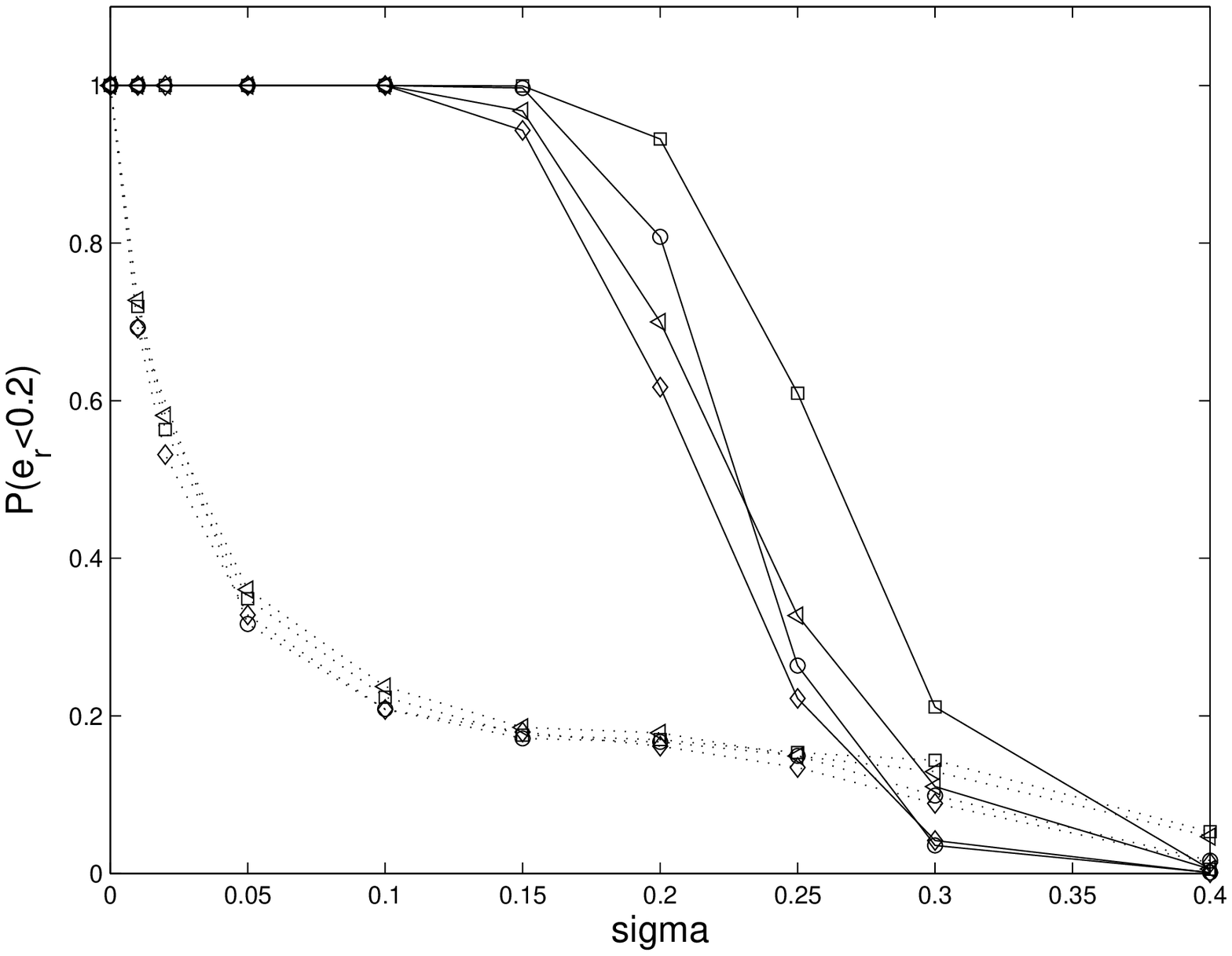}}
\end {subfigure}
\begin{subfigure}{
\includegraphics[height=0.45\linewidth]{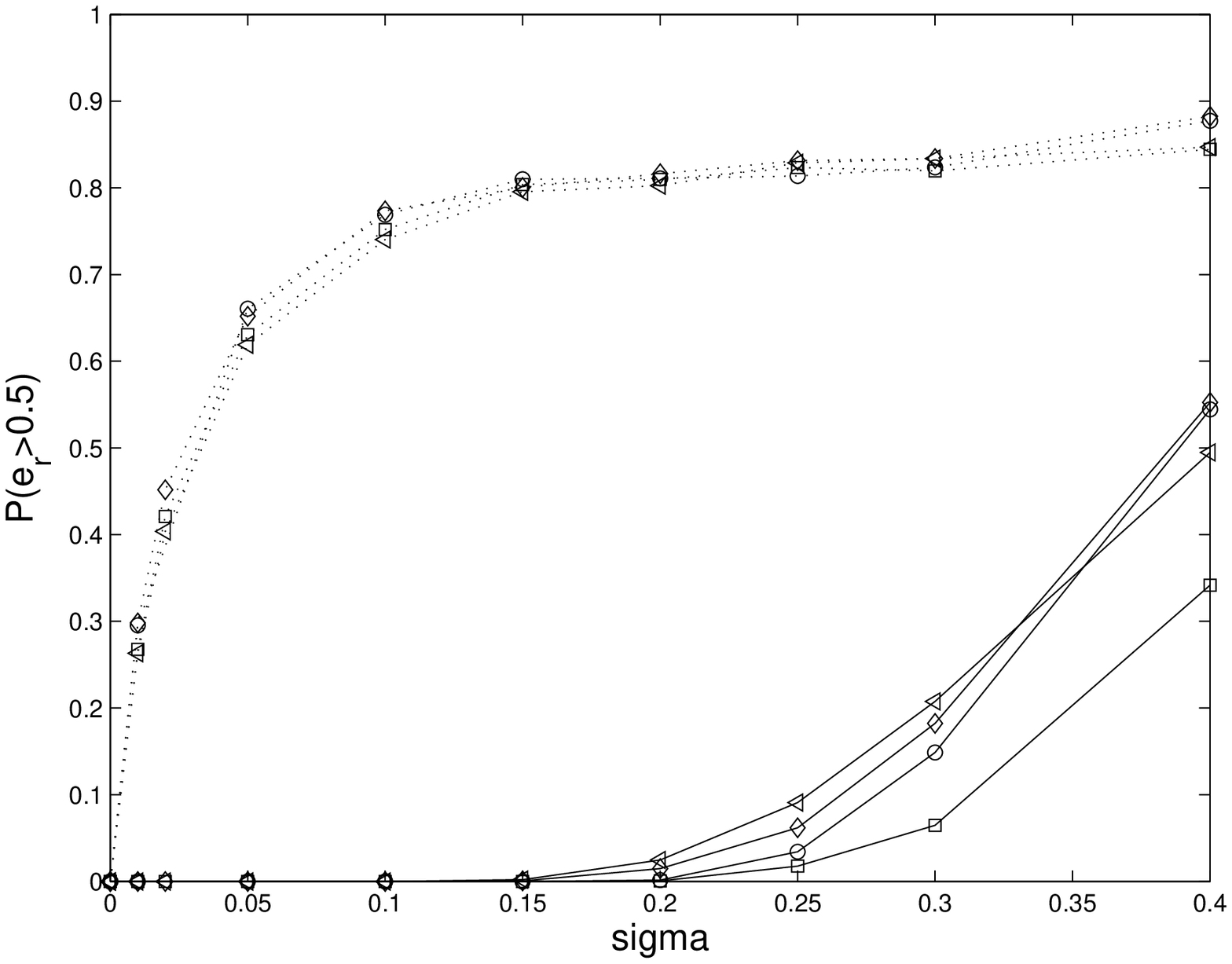}}
\end {subfigure}}
\end{center}}
\caption{performances of the EM and KP algorithms on scenario B for different values of $\sigma$. For each value of $\sigma$ and for each sub-scenario 10000 simulation run have been performed. For each simulation run, 200 observations have been generated. $e_r$ is the maximal distance between the sorted vector of true cluster means and the sorted vector of estimated cluster means. The performance criteria are the probabilities for $e_r$ to be smaller than 0.1 (top), smaller than 0.2 (middle) and greater than 0.5 (bottom).} 
\end{figure}

\begin{figure}{
\begin{center}{
\begin{subfigure}{
\includegraphics[height=0.45\linewidth]{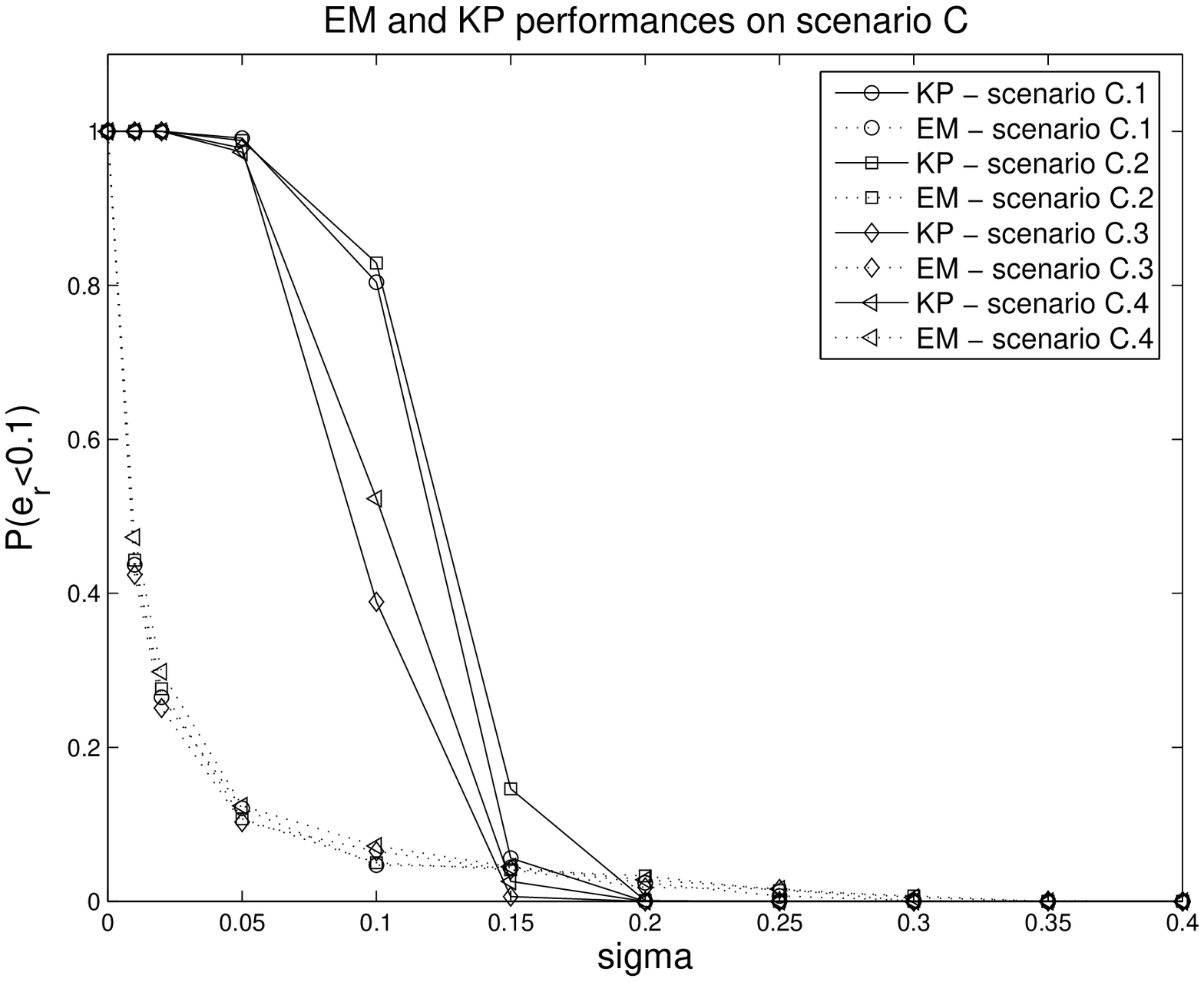}}
\end {subfigure}
\begin{subfigure}{
\includegraphics[height=0.45\linewidth]{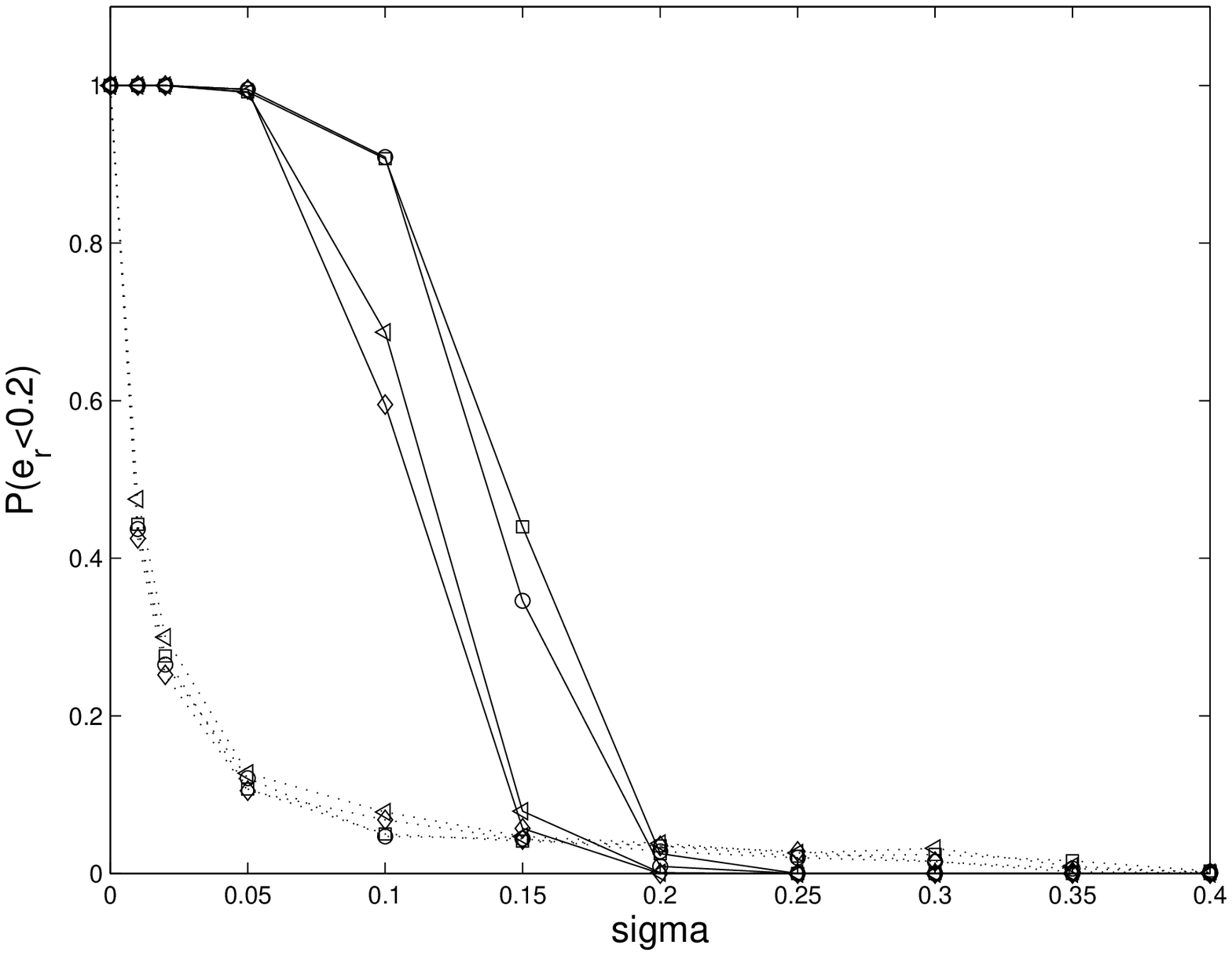}}
\end {subfigure}
\begin{subfigure}{
\includegraphics[height=0.45\linewidth]{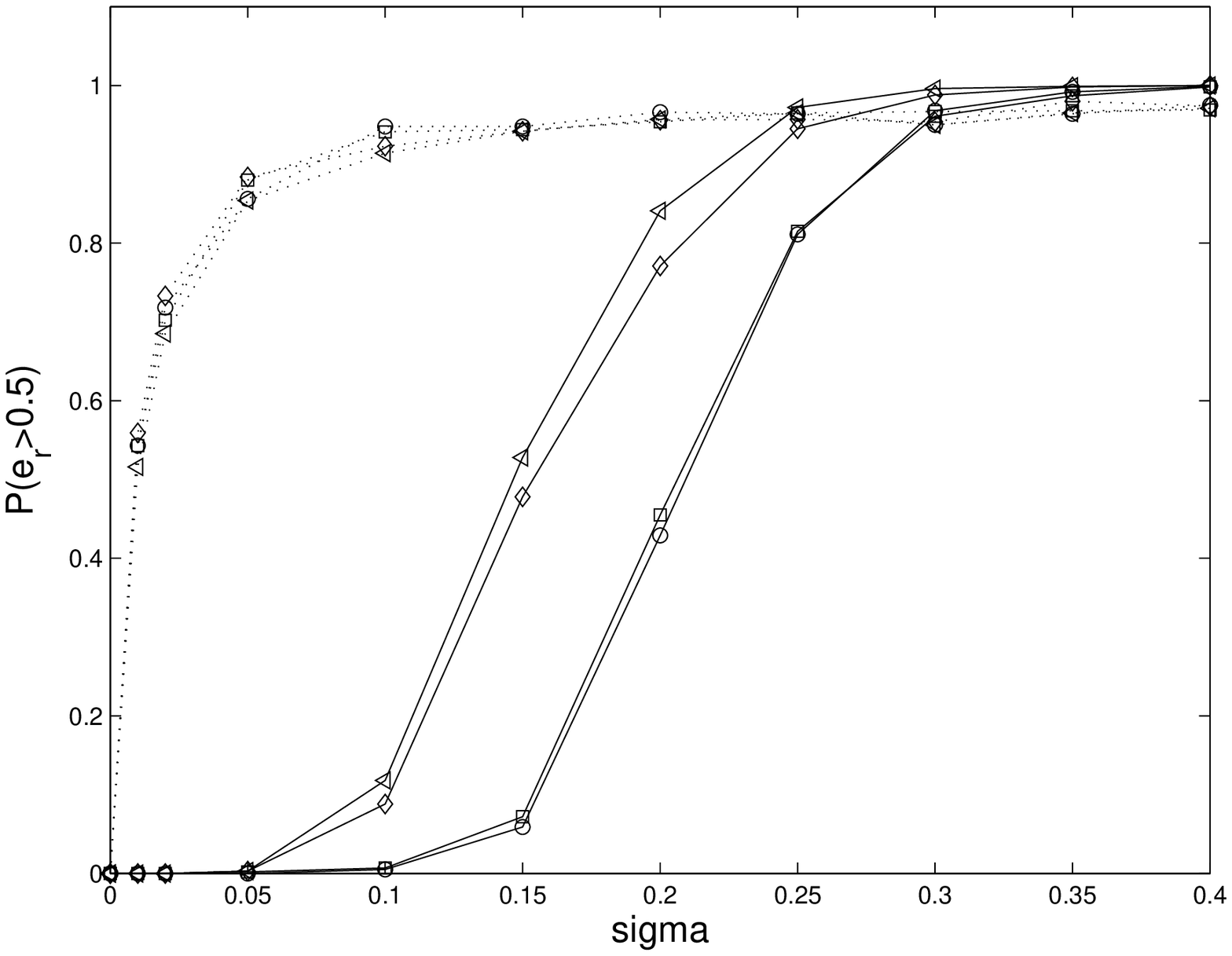}}
\end {subfigure}}
\end{center}}
\caption{performances of the EM and KP algorithms on scenario C for different values of $\sigma$. For each value of $\sigma$ and for each sub-scenario 1000 simulation run have been performed. For each simulation run, 300 observations have been generated. $e_r$ is the maximal distance between the sorted vector of true cluster means and the sorted vector of estimated cluster means. The performance criteria are the probabilities for $e_r$ to be smaller than 0.1 (top), smaller than 0.2 (middle) and greater than 0.5 (bottom).}
\end{figure}

\section{Conclusion}
We have proposed a clusters estimation algorithm for univariate observations when the number of clusters is known. It is based on the minimization of the "KP" criterion we first introduced in (Paul et al. 2006). We have shown that the global minimum of this criterion can be reached with a linear least square minimization followed by a roots finding algorithm. This minimum is used to get a first raw estimation of the cluster means, and a final clustering step enables to recover the cluster means. The proposed method is not iterative, its complexity is in $\text{O}(NK+K^2)$ and it does not require the configuration of any extra parameter. Simulations have illustrated the KP algorithm performances and superiority to the Expectation-Maximization algorithm which can get stuck at a local maximum of the likelihood. We focused on the univariate case and our current researchs deal with the multivariate case. If the observations $\textbf{z}_n$ belong to $\mathbb{R}^d$, if $\{\textbf{x}_k\}_{k\in\{1 \cdots K\}}$ is any set of $K$  vectors of $\mathbb{R}^d$, the KP criterion is now defined as the sum of all the K-terms products $\prod_{k=1}^{K}||\textbf{z}_n-\textbf{x}_k||_{\mathbb{R}^d}^2$. The minima of such criterion and some algorithms to reach them are currently being studied. 

\begin{acknowledgements}
The authors want to thank B. Scherrer, M. Bellanger, P. Tortelier, G. Saporta and J.P.Nakache for their constructive comments that helped in improving this manuscript. 
\end{acknowledgements}



\appendix{}
\section{: non-singularity of \textbf{Z}} 
In appendix A we explain why the matrix $\textbf{Z}$ of size $K \times K$, defined in \eqref{definition_Z} is regular if the number of different observations is greater than $K-1$. $\textbf{Z}$ can be written as the following matrix product:
\begin{equation}
\textbf{Z}=\textbf{V}\textbf{V}^t \notag
\end{equation}
where \textbf{V} is a $K \times N$ Vandermonde Matrix defined by:
\begin{equation}
\textbf{V}\stackrel{\Delta}{=} \left( \textbf{z}_1, \textbf{z}_2, \cdots \textbf{z}_N \right) \notag
\end{equation}
and $\textbf{z}_n$ has been defined in \eqref{definition_zn}. Let us assume that the $K$ first observations are different. The determinant of the $K \times K$ Vandermonde matrix $\left( \textbf{z}_1, \textbf{z}_2, \cdots \textbf{z}_K \right)$ is equal to $\prod\limits_{1 \leq i < j \leq K}{(z_j-z_i)}$, which is different from zero. The rank of $\textbf{V}$ is then equal to $K$, so the rank of \textbf{Z} is equal to $K$ and \textbf{Z} is regular.

\section{: proof of theorem 1}
\noindent In appendix B we prove theorem 1. Let \textbf{$F$} be the function defined by:
\begin{equation}
F: \mathbb{C}^K\rightarrow \mathbb{R}^+: 
\textbf{x} \rightarrow \sum_{n=1}^{N}{ \prod_{k=1}^{K}{||z_n-x_k||_{\mathbb{C}}^2} } \notag
\end{equation}
\noindent The restriction of $F$ to $\mathbb{R}^K$ is the function $J$ since the observations $z_n$ are real:
\begin{equation}
\forall \textbf{x} \in \mathbb{R}^K: \ \ \ F(\textbf{x})=J(\textbf{x})
\label{restrictionRK}
\end{equation}
\noindent Now let $H$ be the function defined by:
\begin{equation}
H: \mathbb{C}^K\rightarrow \mathbb{R}^+: 
\textbf{y} \rightarrow \sum_{n=1}^{N}{\left\|z_n^K-\textbf{z}_n^t\textbf{y}\right\|_{\mathbb{C}}^2} \notag
\end{equation}
\noindent The function $H$ applied to the ESP of a vector $\textbf{x}$ in $\mathbb{C}^k$ is equal to the function $F$ applied to $\textbf{x}$:
\begin{equation}
\forall \textbf{x} \in \mathbb{C}^K: \ \ \ 
F(\textbf{x})=\sum_{n=1}^{N}{ \left\|\prod_{k=1}^{K}{(z_n-x_k)}\right\|_{\mathbb{C}}^2 }
\label{F_SumNormProduct}
\end{equation}
\noindent developping~\eqref{F_SumNormProduct} using definition~\eqref{definition_wk} leads to:
\begin{equation}
\forall \textbf{x} \in \mathbb{C}^K: \ \ \
F(\textbf{x})=\sum_{n=1}^{N}{ \left\|{z_n^K-\sum_{k=1}^{K}{z_n^{K-k}w_k(\textbf{x})}}\right\|_{\mathbb{C}}^2 } \notag
\end{equation}
\noindent including definitions~\eqref{definition_zn} and~\eqref{definition_w}:
\begin{equation}
\forall \textbf{x} \in \mathbb{C}^K: \ \ \
F(\textbf{x})=\sum_{n=1}^{N}{ \left\|{z_n^K-\textbf{z}_n^t\textbf{w}(\textbf{x})}\right\|_{\mathbb{C}}^2 } \notag
\end{equation}
\begin{equation}
\forall \textbf{x} \in \mathbb{C}^K: \ \ \ F(\textbf{x})=H( \textbf{w}(\textbf{x}) )  
\label{FuIsHWu}
\end{equation}
\noindent The global minimum of $H$ is the linear least square solution $\textbf{y}_{min}$ given by:
\begin{equation}
\textbf{y}_{min}=\underset{\textbf{y} \in \mathbb{C}^K}{\text{argmin}}  \left\{\sum_{n=1}^{N}{\left\|z_n^K-\textbf{z}_n^t\textbf{y}\right\|_{\mathbb{C}}^2}\right\}
\label{argminH}
\end{equation}
\noindent developping~\eqref{argminH} using definitions~\eqref{definition_z} and~\eqref{definition_Z} and remembering that the coefficients of $\textbf{Z}$ and $\textbf{z}$ are real:
\begin{equation}
\textbf{y}_{min}=\underset{\textbf{y} \in \mathbb{C}^K}{\text{argmin}}  \left\{\textbf{y}^H\textbf{Z}\textbf{y}-2\text{Re}\{\textbf{y}^H\}\textbf{z} \right\} \notag
\end{equation}
\begin{equation}
\textbf{Z}.\textbf{y}_{min}=\textbf{z}, \ \ \ \textbf{y}_{min} \in \mathbb{R}^K
\label{ymin_obtention}
\end{equation}
\noindent The Hankel matrix $\textbf{Z}$ is regular since the number of different observations is greater than $K-1$ (appendix A). System~\eqref{ymin_obtention} therefore has exactly one solution.  Since $\textbf{Z}$ belongs to $\mathbb{R}^{K \times K}$ and $\textbf{z}$ belongs to $\mathbb{R}^K$, $\textbf{y}_{min}$ belongs to $\mathbb{R}^K$. Now let $\textbf{x}_{min}$=$(x_{1,min},\cdots ,x_{K,min})^t$ be a vector containing, in any order, the $K$ (potentially complex) roots of $q_{\textbf{y}_{min}}(\alpha)$. One can show that the following holds: \\
\noindent \hspace*{1cm}(i) $\textbf{x}_{min}$ is a global minimum of $F$ \\
\hspace*{1cm}(ii) $\textbf{x}_{min} \in \mathbb{R}^K$ \\
\hspace*{1cm}(iii) $\textbf{x}_{min}$ is a global minimum of $J$  \\
\noindent Property (i) is a direct consequence of~\eqref{FuIsHWu}:
\begin{equation}
\forall \textbf{x} \in \mathbb{C}^K: \ \ \ F(\textbf{x})=H(\textbf{w}(\textbf{x})) \notag
\end{equation}
\begin{equation}
\forall \textbf{x} \in \mathbb{C}^K: \ \ \ F(\textbf{x}) \geq \text{min} \left\{H\right\} \notag
\end{equation}
\begin{equation}
\forall \textbf{x} \in \mathbb{C}^K: \ \ \ F(\textbf{x}) \geq H(\textbf{y}_{min}) \notag
\end{equation}
\noindent According to (\ref{roots_coefficients}), $\textbf{y}_{min}=\textbf{w}(\textbf{x}_{min})$ and we have:
\begin{equation}
\forall \textbf{x} \in \mathbb{C}^K: \ \ \ F(\textbf{x}) \geq H(\textbf{w}(\textbf{x}_{min})) \notag
\end{equation}
\begin{equation}
\forall \textbf{x} \in \mathbb{C}^K: \ \ \ F(\textbf{x}) \geq F(\textbf{x}_{min}) \notag
\end{equation}
\noindent which proves (i). Property (ii) can be shown by contradiction: if $\textbf{x}_{min}$ does not belong to $\mathbb{R}^K$, then for one of the $x_{k,min}$ we have $x_{k,min} \neq \text{Re}\{x_{k,min}\}$ and, since all the observations $z_n$ are real:
\begin{equation} 
\forall n \in \{1,\cdots,N\}: \ \ \
\left\|z_n-x_{k,min}\right\|_{\mathbb{C}} > \left\|z_n-\text{Re}\{x_{k,min}\}\right\|_{\mathbb{C}} \notag
\end{equation}
\noindent which leads to:
\begin{equation}
F(\textbf{x}_{min})>F(\text{Re}\{\textbf{x}_{min}\}) \notag
\end{equation}
\noindent This is impossible since $\textbf{x}_{min}$ is a global minimum of $F$. This proves property (ii). We finally have to prove (iii): since $\textbf{x}_{min} \in \mathbb{R}^K$ we have, using~\eqref{restrictionRK}: 
\begin{equation}
F(\textbf{x}_{min})=J(\textbf{x}_{min}) 
\label{FuminIsJumin}
\end{equation}
\noindent Furthermore, according to~\eqref{restrictionRK}:
\begin{equation}
\forall \textbf{x} \in \mathbb{R}^K: \ \ \ J(\textbf{x})=F(\textbf{x}) \notag
\end{equation}
\begin{equation} 
\forall \textbf{x} \in \mathbb{R}^K: \ \ \ J(\textbf{x}) \geq \text{min}\{F\} \notag
\end{equation}
\noindent then, according to property (i):
\begin{equation}
\forall \textbf{x} \in \mathbb{R}^K: \ \ \ J(\textbf{x}) \geq F(\textbf{x}_{min}) \notag
\end{equation}
\noindent using~\eqref{FuminIsJumin}:
\begin{equation}
\forall \textbf{x} \in \mathbb{R}^K: \ \ \ J(\textbf{x}) \geq J(\textbf{x}_{min}) \notag
\end{equation}
\noindent which proves (iii). Properties (ii) and (iii) directly lead to theorem 1.

\begin{thebibliography}{}

\addtolength{\leftmargin}{0.2in}
\setlength{\itemindent}{-0.2in}

\bibitem{berkhin}
Berkin P (2006) A Survey of clustering data mining techniques. Grouping Multidimensional Data: Recent Advances in Clustering, Ed. J. Kogan and C. Nicholas and M. Teboulle, Springer, pp. 25-71 
\bibitem{bradley}	
Bradley P S, Fayyad U M (1998) Refining initial points for K-means clustering. Proc. of the 15th Int. Conf. on Machine Learning, San-Fransisco, Morgan Kaufmann, pp.  91-99 
\bibitem{celeux}	
Celeux G, Chauveau D, Diebolt J (1995) On stochastic version of the EM algorithm. INRIA research report no 2514, available: http://www.inria.fr/rrrt/rr-2514.html 
\bibitem{dempster}	
Dempster A, Laird N, Rubin D (1977) Maximum likelihood from incomplete data via the EM algorithm. Journal of the Royal Statistical Society, B. 39, pp. 1-38
\bibitem{fisher} 
Fisher W D (1958) On grouping for maximum homogeneity. Journal of the American Statistical Association, Vol. 53, No. 284, pp. 789-798 
\bibitem{fitzgibbon} 
Fitzgibbon L J, Allison L, Dowe D L (2000) Minimum message length grouping of ordered data. Algorithmic Learning Theory, 11th International Conference, ALT 2000, Sydney, Australia 
\bibitem{hartigan}
Hartigan J, Wong M (1979) A k-means clustering algorithm, Journal of Applied Statistics, vol 28, pp. 100-108 
\bibitem{krishna} 
Krishna K, Narasimha Murty M (1999) Genetic K-Means Algorithm, IEEE Transactions on Systems, Man, and Cybernetics - Part B: Cybernetics, Vol. 29, No. 3 
\bibitem{lindsay2}	
Lindsay B, Furman D (1994) Measuring the relative effectiveness of moment estimators as starting values in maximizing likelihoods. Computational Statistics and Data Analysis, Volume 17, Issue 5, pp. 493-507 
\bibitem{mclachlan} 
McLachlan G, Peel D (2000) Finite Mixture Models. Wiley Series in probability and statistics, John Wiley and Sons
\bibitem{parzen} 
Parzen E (1962) On estimation of a probability density function and mode. Annals of Mathematicals Statistics 33, pp. 1065-1076 
\bibitem{norsig} 
Paul N, Terre M, Fety L (2006) The k-product criterion for gaussian mixture estimation. 7th Nordic Signal Processing Symposium,  Reykjavik, Iceland 
\bibitem{pernkopf}
Pernkopf F, Bouchaffra D (2005) Genetic-based EM algorithm for learning gaussian mixture models, IEEE Transactions On Pattern Analysis and Machine Intelligence, Vol. 27, No. 8 
\bibitem{xu} 
Xu R, Wunsch II D (2005) Survey of Clustering Algorithms. IEEE Transactions On Neural Networks, vol. 16, No. 3, pp. 645-676 
\end{thebibliography}
\end{document}